# Geotagged tweets to inform a spatial interaction model: a case study of museums

Robin Lovelace, Nick Malleson, Kirk Harland and Mark Birkin

*School of Geography, University of Leeds, Leeds, UK*

R.Lovelace@leeds.ac.uk

# Geotagged tweets to inform a spatial interaction model: a case study of museums

This paper explores the potential of volunteered geographical information from social media for informing geographical models of behavior, based on a case study of museums in Yorkshire, UK. A spatial interaction model of visitors to 15 museums from 179 administrative zones is constructed to test this potential. The main input dataset comprises geo-tagged messages harvested using the Twitter Streaming Application Programming Interface (API), filtered, analyzed and aggregated to allow direct comparison with the model's output. Comparison between model output and tweet information allowed the calibration of model parameters to optimize the fit between flows to museums inferred from tweets and flow matrices generated by the spatial interaction model. We conclude that volunteered geographic information from social media sites have great potential for informing geographical models of behavior, especially if the volume of geo-tagged social media messages continues to increase. However, we caution that volunteered geographical information from social media has some major limitations so should be used only as a supplement to more consistent data sources or when official datasets are unavailable.

**Introduction**

Much has been written about the potential of volunteered geographical information (VGI) for informing geographic research projects (1). Geo-coded social media data provided by a diffuse network of self selected users presents a rich yet challenging data source (2) that could, if filtered and interpreted with sufficient care, provide useful inputs into many areas of geographic research (3). In this paper we demonstrate and discuss the potential of VGI as an input into geo-science using a case study of a spatial interaction model (SIM) calibrated using harvested Twitter data.

SIMs have a long history in academic and commercial applications of geo-science. Their utility is demonstrated by persistent use in a range of fields including retail planning and store location optimization (4, 5), transport planning (6) and the investigation of migration (7, 8). To introduce the SIM concepts and illustrate the kind of problem that the technique tackles, let us consider a concrete case study in some detail rather than to attempt an exhaustive review on the large body of literature on the subject. A recent study investigating flows of aggregates – rock-based material used in the construction industry – by Zuo et al. (9) serves this purpose well.

*Spatial interaction models*

SIMs are most frequently applied to estimate the flow rate ($T_{ij}$) between a number of origins (i) and destinations (j), as a function of the distance between them and their characteristics. The output is usually a *flow matrix* with row names corresponding to origins and column headings to destination. The SIM is used to estimate the cell values of the flow matrix. The classic version of a SIM, with the output being the origin-destination flow matrix T, is presented in equation 1 below:

$$T_{ij} = A_i B_j O_i D_j f(d_{ij}) \qquad (1)$$

where $O_i$ and $D_j$ represent each origin and destination (or, in more complex models, some function of their characteristics); $A_i$ and $B_j$ are weights associated with the origins and destinations respectively; and $d_{ij}$ is the distance or transport costs that separate them in geographical space.

As Wilson (10) demonstrates, the treatment of the weights ($A_i$ and $B_j$) is of crucial importance in the modeling process. Without restriction on either of these terms (or when they are absent) the resulting model is considered 'unconstrained'. Alternatively the weights can be used as 'balancing factors' to maintain consistency in either the origin totals or destination totals or both simultaneously, in the case of a doubly constrained model. In the case of the paper on flows of rocky aggregates for the construction industry (9), the origins are the quarries from which materials are mined and the destinations are the districts to which the materials are delivered. Balancing factors are employed to constrain to known aggregate flows from each quarry and to each district. Hence the model was specified with the following functional form (9):

$$T_{ij} = A_i B_j O_i D_j d_{ij}^{-\beta} \qquad (2)$$

$$A_i = 1 \Big/ \sum_j B_j D_j d_{ij}^{-\beta} \qquad (2A)$$

$$B_j = 1 \Big/ \sum_i A_i O_i d_{ij}^{-\beta} \qquad (2B)$$

where $A_i$ and $B_j$ are the balancing factors, $O_i$ and $D_j$ origins and destinations and $\beta$ is the distance decay parameter, to be calibrated. The beta ($\beta$) value is crucial in SIMs as it determines the distance decay effect between origins and destinations and subsequently the extent to which flows between neighboring origins and destinations are favored over longer distance flows. Much research has focused on the variability of $\beta$ values and the functional form of distance-decay $f(d_{ij})$ (11, 12). In this paper we also follow Wilson (10, 13) in choosing an exponential decay:

$$f(d_{ij}) = \exp(-\beta\, d_{ij}) \qquad (3)$$

The destination zones are represented as points, allowing precise distances (d) to be calculated. Final refinements to the SIM included additional factors such as the proximity of transport links to each quarry, once complete parameter values were estimated through *calibration* (9). Calibration involves iteratively estimating parameters to optimize the fit between observed and modeled flows; in the model presented here Twitter data is used to calibrate β in equation 3, and explore other parameters that could be optimized.

In the model of museum visits presented here, the configuration of the SIM is slightly different to that presented by Zuo et al. in (9): the origins are diffuse administrative zones and the destinations are a limited number of unevenly distributed points – the museums – located in and around the English cities of Leeds and Bradford. A simplified representation of this model set-up, for 10 origin zones and 2 museums, based on real locations, is illustrated in Figure 1.

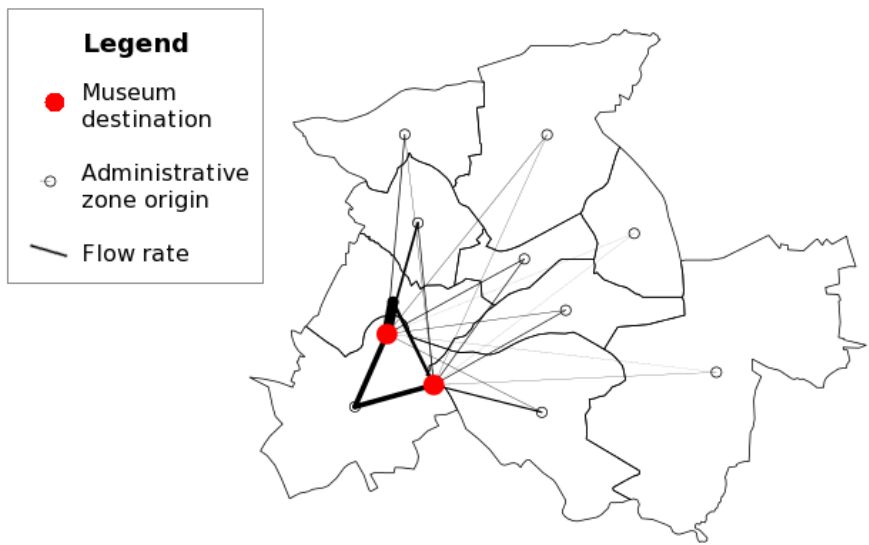

Figure 1. Simplified illustration of an unconstrained spatial interaction mode (SIM), based on 10 ward-level administrative zones and two well-known museums in Leeds. The flow of visitors to each museum is illustrated by the thickness of each straight line.

*Geotagged social media to inform a spatial interaction model*

In this paper we show how the SIM structure described above can be applied, to explore movement patterns of people visiting museums. The key innovation is the use of VGI, taken from Twitter, to inform the model. Methods and external data to effectively calibrate SIMs have been a long-running concern for researchers (5, 14, 15). Critically, we demonstrate that social media data can aid the construction and calibration of SIMs, enabling the inclusion of variables not available from traditional sources. Methods of harnessing geo-referenced social media data for flow model calibration are also demonstrated, highlighting the need for input data that is sufficiently large and rich for the purpose. We developed the methods in R, a free and open source statistical programming language that has already been used for calibrating SIMs (16).

While SIMs have occasionally been applied to the related phenomena of journey-to-work flows (17, 18) and tourist travel (19, 20), these personal travel models tend to assume that patterns are regular and cyclical – in a sense a long-term equilibrium. The reality is much more varied – recent research using GPS tracking devices shows that human behavior is complicated, with seemingly stochastic variability in observed timings and spatial distributions of everyday activity patterns (21, 22). One could also speculate that, with increased communication between people and potential for fragmentation of travel planning sources during the on-going digital revolution (23, 24), the tendency is towards greater complexity. Time-series datasets are generally too scarce to test such hypotheses at present but the general point, that personal travel is often not regular temporally or predictable spatially is widely accepted (25). Such unpredictability and 'messiness' are not well captured in much transport research (see (26) for a discussion of this issue from the perspective of traffic safety modeling), as illustrated by the omission of the following phenomena from most contemporary transport models:

- Multi-purpose journeys, the tendency to chain together trips of different types. The trip chain home-school-work-shop-work-gym-bar-home, for example, may be represented as a simple home-work trip.
- Seasonal, weekly and diurnal variation in the timing and frequency of trips – for example due to holidays, 'flexi-time' work contracts, and changing consumer habits.
- Up-to-date information about the state of the travel network due to improved internet communications in combination with the rapid uptake of smartphones increasingly influences the timing and route choice of trips. Real-time traffic information from Google or directly to in-car satellite navigation systems and the current state of bike hire hubs are examples of factors that could influence timings, mode and route choice.

The above points are especially relevant to irregular trip demands (e.g. shopping, gym, leisure): these are less predictable than regular trips such as commuting, and are therefore challenging to model. The study of museum visits in this paper is an excellent example of such irregular trip-making behavior. One difficulty associated with analyzing the extent of variability in travel patterns is that the data has been difficult and expensive to acquire - most travel diary and other official data collection techniques tend to assume regular trips and mask variability. This is starting to change as part of the 'Big Data revolution': crowd-sourced, social media and passively harvested datasets are being collected on a vast scale. The academic sector has been slow to take advantage of these new datasets (27) despite widespread acknowledgment of their potential for developing our understanding of travel patterns and behaviors. One barrier to uptake is data quality: tweets, for example, contain a maximum of only 140 characters and are not constrained in any other way, posing major challenges for interpretation. Another obstacle is that social media users are a self-selecting sample, unlikely to be representative of the wider population.

One could argue that the most prolific online communicators are wholly *unrepresentative*, constituting a small minority of the population with an unusual attachment to digital device – the geo-referenced tweets used in this paper, for example, constitute only around 2% of Twitter information output and cannot be assumed to be representative of all tweets, let alone of society as a whole (2). Academics have rightly been cautious of these issues, but we argue that much more use can be made of the vast stores of volunteered geographical information that are available, especially in the data deprived area of modeling complex irregular travel patterns.

*Research overview*

This paper examines ways of tackling this research gap with an exploratory example of museum visits and communications about museums, utilizing Twitter data as a source of information about attitudes towards and spatial patterns of museum visits. These social media data were captured by harvesting tweets in the case study area surrounding Leeds and Bradford, UK, over an 18 month period. The geographic data on museums was collected from Open Street Map. The following section of the paper describes the input data in more detail. Various filters were used to extract subsets of 'museum tweets' from the parent data set and the properties of this subset are described and analyzed. The steps which have been taken to infer the residence of the 'tweeter', the location of the museum or exhibition and other relevant places e.g. the source of the message are also presented.

Section 3 outlines how a simple unconstrained SIM was used to estimate the flows between administrative zones and museums and the steps taken to select initial model parameters during calibration. Refinements to the model consider variations in the attractiveness of different exhibits, visitor demographics and trip-making propensities. The final section reviews and discusses the findings and presents concluding comments.

Although this research project engages directly with the concept of 'big data' the extract analysed here is actually rather limited; avenues for extending the sample size are being considered. Potential applications for a model of this kind are widespread, for example it could be used to examine the effectiveness of advertising for specific shows, or to track public sentiment towards and appreciation of diverse exhibitions amongst different audiences. More interesting from an academic view perhaps is how this begins to illustrate novel interaction patterns in a city, which are not only interesting in their own right but perhaps even more so as individual elements in an increasingly complex web of daily urban movements.

**Data and Methods: harvesting and filtering the tweets**

*The museum data*

The museums in the case study area were identified using a spatially bounded search of Open Street Map (OSM) data. The entirety of UK OSM data was first download in the compressed Protocolbuffer Binary Format (.pbf). To reduce processing time, this file was clipped to the study area.[1] After transferring the data into a spatial database and then a GIS, the attribute table was analyzed. . Both 'name' and 'tourist' columns were used to search for the keyword 'museum', without case sensitivity. All named museums were also classified as museums, so the tourist attribute variable was used for identification.[2]

In total there were 7 point objects and 13 polygon objects tagged as "museum" in the column "tourism". However, one of the points[3] duplicated a pre-existing polygon of the same building and so was deleted. Bradford Industrial Museum, in the north-eastern outskirts of the city, is composed of multiple buildings and was replicated 5 times. Once these issues had been resolved, the total number of museums in the study area was 15. These were all converted into point data for consistency (figure 2).

---

[1] To clip the .pbf file the command-line OSM conversion tool 'osmconvert' was used. The following command was applied to the nationwide data, resulting in a 5-fold reduction in filesize: osmconvert leedstw.osm.pbf > leedstw.osm.

[2] The attributes of OSM elements, including 'name' and 'tourism' are described here: http://wiki.openstreetmap.org/wiki/Map_Features

[3] This point was Bolling Hall Museum, on the Southern outskirts of Bradford. Note in figure 2 that this museum is misspelt in the map, underlying data quality issues associated with OSM data.

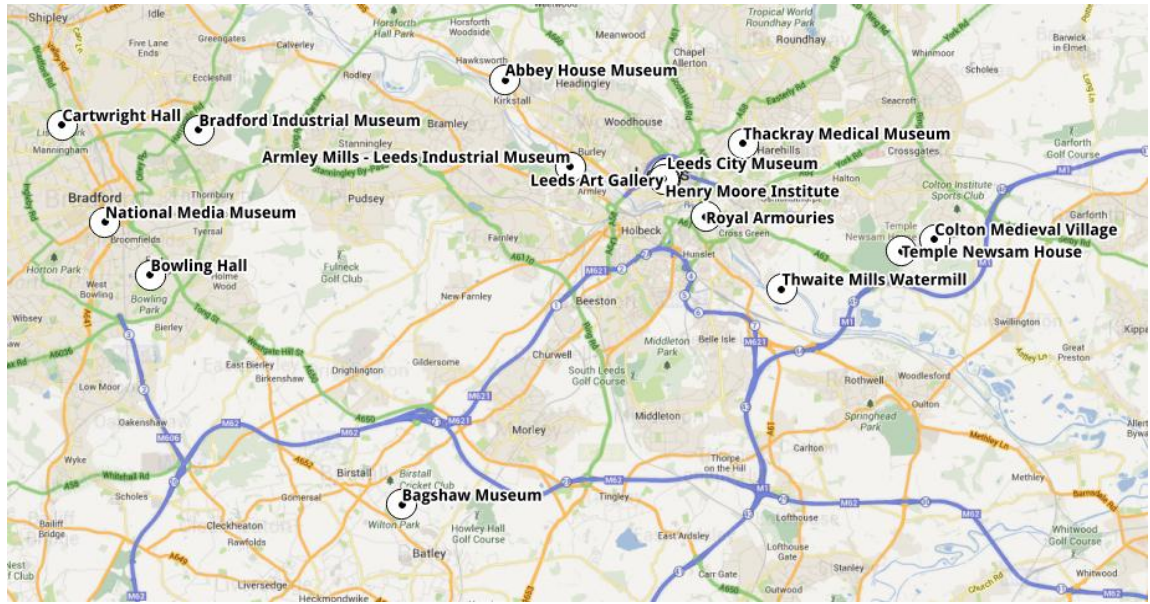

Figure 2. Locations of the 15 museums used in the case study. Basemap: Google.

The Twitter data were collected using the Twitter Streaming API. This is a service that provides limited access to public messages posted to Twitter. Data were collected during 445 days between 2011-06-22 and 2012-09-09. The full dataset consisted of 992,423 tweets originating in the Leeds area. Each record in the dataset represents one tweet and includes a timestamp for the generation time, a user id allowing tweets from the same account to be linked together and geographical coordinates of the location where the tweet originated. Only messages that have associated geographical coordinates have been included in the analysis – recent estimates suggest that this represents 1-2% of all messages (2). Such messages are commonly created using mobile devices by users who have explicitly opted to publish their present location.

Using the data manipulation functions in the MySQL database management system a cursory analysis of the collected dataset was undertaken. Most user accounts showed tweeting activity of less than 1,000 during the period of study. However, a small number of accounts, 274, had activity much higher than 1,000 tweets. Further investigation revealed 12 accounts that were both prolific users and geographically static. Further examination of the linguistic structure within the tweeted text identified repetitive patterns representing weather stations, car sales promotions, traffic warnings and advertising for online magazines. The 12 accounts issuing automated information were removed from subsequent analysis. The remaining dataset composed 958,339 tweets over 27,999 active user accounts.

The text from each tweet was decomposed into a series of words. The rule used to identify words in this analysis is a set of characters encapsulated by either blank space or a combination of blank space and a punctuation character such as a full stop, comma, exclamation mark, question mark, colon or semi-colon. Words consisting of one or two characters were excluded from further analysis leaving 11,505,719 words forming the 958,339 tweets (28).

*Filtering the tweets*

The next stage in the analysis was to filter the tweets to identify those relating to museum visits allowing the investigation of the spatial behavior of museum visitors. Two main options for filtering a predefined study area are to apply a *spatial* or a *semantic* filter. The former is usually simply implemented as a 'clip' function to remove any information outside the areas of interest, but can also be applied by altering the probability of selection or rank of observations based on their proximity to the location of interest (29). Semantic filters interpret the user's meaning from character strings within the text of their message. This type of textual filter can also be constructed at different levels of complexity, from simple keyword searches - all Tweets containing "museum", for example - to more complex techniques (30). More sophisticated searches include 'fuzzy matching' to account for misspellings or synonyms (31) and the classification of tweet content into semantic categories to extract greater meaning from the combination of words used (32). For example by allocating facets, or meaning categories, using techniques of Natural Language Processing (NLP) (33). Semantic searches can go far beyond simple keyword searches, and it has been found that this can yield benefits to users looking for context-specific information (34). Below we illustrate methods to implement both types of search to select tweets about museums.

The *spatial filter* in this context would select only those tweets recorded within or very close to known museums. Below we use the Royal Armouries, a national museum of history and armaments that moved to Leeds in 1996 (35), to illustrate the method. The first stage was to identify its geographical coordinates, using an online search. We used the search engine 'www.geonames.org', although any one out of a number of sites yielded the same lat/long coordinates for its centroid: 50.8607 / -1.1389. A straightforward spatial filter would select all tweets with a certain distance (e.g. 200 m) of this location, to capture all tweets sent in and direct surrounding the building. This is clearly an inadequate method because museums are rarely circular buildings and their size changes greatly from one to the next (see figure 3). A *floor-plan* of the museums under investigation is required so a more accurate spatial filter can be created. This data is available from a variety of local sources including the local Council or the Estate Manager at the museum itself. Building floor plan polygons are also available at the national level for the UK from the Ordnance Survey's MasterMap product, however this dataset is large and requires a license for access making access problematic.[4] For reproducibility, ease of access and (albeit patchy) international coverage, we decided to use Open Street Map (OSM), a crowd sourced worldwide map from which the raw data is publicly available.

---

[4] MasterMap data can be downloaded for free under an academic license, but only in 10 by 10 km chunks. The size of these 100 km2 chunks is large, 160 Mb for the area directly surrounding Leeds.

To access OSM vector data, and not just the raster data that is displayed on the OSM website, we used osmar, an R package that connects to OSM's API. Based on the aforementioned grid reference, the following code was used to download vector data containing the floor-plan of the Royal Armouries Museum and save it as an object called 'arm':

```
bb <- center_bbox(-1.5323, 53.7919, 60, 60)

arm <- get_osm(bb, source=src)
```

This resulted in the polygon shown in figure 3. After the polygon was converted into a conventional R spatial object and its coordinates were transformed into a local coordinate system (OSGB 1936, EPSG code 27700), the area was calculated and the true centroid of the building in the new coordinate system was recorded. 99 tweets were recorded from within a 10 metre buffer of the museum's floor-plan. Of these only one was also present in both samples, and the message appeared to be unrelated to either the Royal Armouries, or museums in general.[5]

---

5 "Totally just tried and failed to wangle our way into a food exhibition at Clarence Dock in Leeds. Who knew they'd CHECK!"

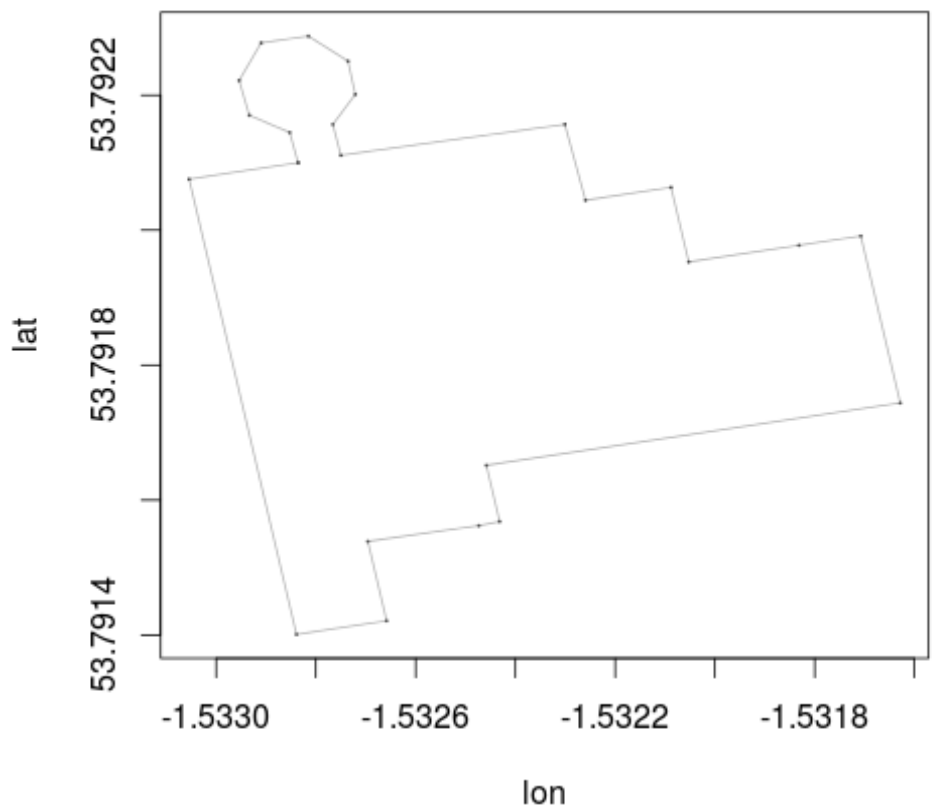

Figure 3. Floor plan of the Royal Armouries Museum, obtained from raw Open Street Map data.

Although this methodology succeeds at collecting tweets geographically associated with museums, it largely fails at returning tweets that are linked to museums *semantically*: just because a tweet is sent from near a museum, does not mean it is about a museum. In addition, museums are often closed spaces so it is not clear whether or not an accurate GPS position can be established by the mobile device before transmitting a message. Hence tweets from inside museums might not be accurately positioned and appear outside a spatial filter. Visitors communicating with their friends and people who are simply near the museum coincidentally (e.g. to use a café associated with the building) seem to constitute the majority of these tweets: only 1 of the 99 tweets from

the Royal Armouries Museum could be identified with any direct semantic link to the weapons on display.[6]

To overcome these issues a semantic filter was used. This means scanning the text contents of the tweets and selecting those that contain certain keywords. There are many strategies that can be employed to increase the proportion of relevant tweets selected whilst simultaneously reducing unwanted 'false positives' – tweets that are not relevant to the topic under investigation. Here, all tweets containing one or more of the words 'museum', 'gallery', 'exhibition' and 'exhibit' were selected. This filter produced a subset of 1,553 tweets from 684 individual user accounts. These were clustered around the urban centers of Leeds and Bradford (figure 5).

Selecting an appropriate level of investigation for the origin areas was critical: if the areas were too small, there would be many zones containing no social media or socio-demographic data; too large and SIM would contain insufficient geographical detail to be useful. The number of zones containing semantically filtered museum tweets was assessed using a spatial intersect query[7]. The results for a number of administrative scales are presented in Table 1. Based on these numbers, and the shapes of the administrative zones, Census wards were selected as the administrative zone with the most appropriate balance of geographical detail and spatial aggregation.

---

[6] The tweet included the text "Joust time!" and a link to a photo. Another tweet was discovered from a 20 m buffer, that was definitely associated with a museum visit: "Oliver with his new gun" and a link to a photo of a child next to an old cannon.

[7] In this instance the 'select by location' function in QGIS was used.

Table 1. Geographic levels considered for use as the spatial units in the SI model.[8]

| Level | N. zones | N. zones with tweets | N. Tweets/zone | Av. Pop. |
|---|---|---|---|---|
| OA | 4000 | 269 | 3.4 | 300 |
| MSOA | 191 | 121 | 7.7 | 6300 |
| Census ward | 65 | 64 | 14.5 | 18500 |
| Constituencies | 16 | 16 | 58.0 | 75000 |
| Local Authority | 6 | 6 | 154.7 | 200000 |
| TTW Zone | 5 | 5 | 185.6 | 240000 |

In most cases it can be assumed that people do not send tweets from their home locations about museums. A user's home location was assumed to be where the highest number of tweets occurred for a user account. Using a simple aggregate query over location (generalized to 100 meters) and the unique identification number for account holders, the locations with the highest tweeting activity were identified. The distribution of these favorite tweet locations, henceforth referred to as 'home locations' (although it is acknowledged that these may not always correspond with users' true home locations) is substantially larger, incorporating an area roughly double the size of the 'museum tweets'. The study area was determined by this larger spatial extent, specifically all 7 Local Authorities illustrated in Figure 5.

---

[8]See here for definitions and descriptions of UK administrative zones:

http://www.ons.gov.uk/ons/guide-method/geography/beginner-s-guide/index.html

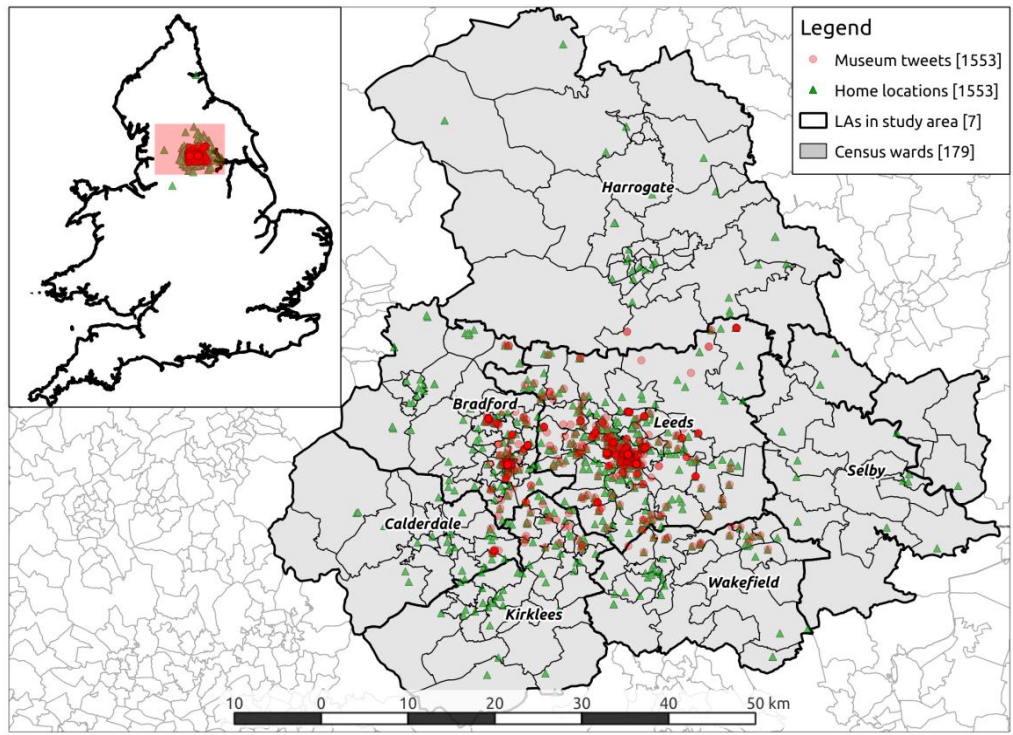

Figure 4. Overview of the geographical distribution of the semantically filtered museum tweets (red dots) and home locations (green triangles). The shade of points corresponds to density, illustrating high densities in Leeds (centre) and Bradford (to the west).

Analysis of the most frequent tweeters (users whose ids were associated with more than 10 museum tweets) led to the discovery of repeated tweets. R's 'unique' function was used to ascertain the proportion of repeated tweets: 27% for all tweets, increasing to 31% for the most frequent tweeters. Additional analysis found that even for tweets with unique text contents, in many cases the only difference between one tweet and the next was the html code associated with the tweet. In every case this was found to be due to

Foursquare, which automatically sends geographical tweets from certain locations (36).[9] It is possible that some of these 'auto check-in' messages triggered simply due to proximity to a museum rather than genuine interest and it is clear that these tweets are not of the same value as those who deliberately composed messages about museums and could potentially be misleading. The foursquare tweets were identified and removed leaving 928 unique tweets remaining. These were the data points used to inform the SIM.

*A Spatial Interaction model of museum visits*

In the absence of official visit data from museums in the study area, a SIM cannot be precisely constrained in terms of flows by destination. Data on the flows emanating from each origin is scarcer still. Another issue to consider in this sub-national case study is that not all museum destinations are considered: only those in Leeds and Bradford Local Authorities (see figure 5) – although Leeds is somewhat of a regional center, many residents, especially those located far from local museums will travel further afield.[10] For these reasons an unconstrained SIM was used (11). Building on equation 1, the flows ($T_{ij}$) were estimated as follows:

$$T_{ij} = Inc_i P_i W_j \exp(-\beta\, d_{ij}) \tag{4}$$

---

[9] This capability, and how to disable it, is further described online:

http://aboutfoursquare.com/do-yourself-a-favor-and-stop-sending-every-checkin-to-twitter-and-facebook/

Where Inci is the income-adjusted demand for museum trips per unit population (P) in each zone and β the distance-decay parameter introduced earlier. $W_j$ is the 'attractiveness' of museum j, calculated as a composite of factors.

*Estimating museum attractiveness*

It is clear that a wide range of factors influence how 'attractive' museums are. Some people will travel to museums because they have a specialist interest in the objects on display; others will seek out museums based on family considerations; whether there are facilities for children or cooked food, for example. Quantitatively, attractiveness can be seen as a coefficient (W) that is a function of various contributing factors:

$$W_j = X1^{a1} X2^{a2} ... Xn^{an} \tag{5}$$

where X is a vector of factors related to museum attractiveness and a1, a2 etc. are coefficients representing how important each is considered (13). This allows the attractiveness of each museum to be determined by any number of factors. In our model we use only two factors: floor area and number of mentions in the media based on searches in Google News. The former is an indication of capacity (and could be further refined, for example by including number of floors); the latter is an indication of how well-known the museum is. There is great variability in these proxies of attractiveness, as illustrated in Table 2. It is clear that the National Media Museum and the Royal Armouries are by far the most well-known. We used a modified version of equation (5) to determine attractiveness:

$$W_j = 0.5 FA^{0.5} + 0.3 MM^{0.5} \tag{6}$$

where FA is floor area and MM is number of mentions in the media, normalized to one.

Table 2: Museum characteristics and proxies of attractiveness.

| Museum | Tweet count | Mean home-museum dist. (km) | Average tweet-museum dist. (km) | Museum floor plan (m2) | News Mentions |
|---|---|---|---|---|---|
| Abbey House Museum | 8 | 2.9 | 132 | 1072 | 2 |
| Armley Mills | 55 | 3.5 | 194 | 2734 | 2 |
| Bradford Industrial Museum | 11 | 5.6 | 110 | 1382 | 1 |
| Cartwright Hall | 2 | 8.5 | 95 | 1519 | 4 |
| Henry Moore Institute | 25 | 6.6 | 86 | 562 | 5 |
| Leeds Art Gallery | 93 | 5.5 | 115 | 1322 | 8 |
| Leeds City Museum | 102 | 5.2 | 130 | 1731 | 7 |
| National Media Museum | 288 | 8.5 | 131 | 3211 | 252 |
| Royal Armouries | 154 | 6.4 | 134 | 5180 | 36 |
| Thackray Medical Museum | 18 | 13.7 | 136 | 1790 | 5 |

*Estimating demand*

Demand for museum visits can be seen as a composite of the total population of an area and the inhabitants' propensity to visit museums. In the model a weighting factor (Inc) was used to weight the population of each ward:

$$Inc_i = 0.1 + Arts_i + 0.03Ea_i \qquad (7)$$

Where Arts is the proportion of individuals who frequent 'fine arts' establishments and Ea is a proxy of average earnings, in each ward area. These data are generated from a substantial lifestyle survey of the UK population (37). Unsurprisingly, Arts and Ea variables are correlated, albeit weakly (r = 0.20) – higher earners more frequently visit fine arts attractions. As with W, Inc was divided by its mean to set the mean equal to one, allowing direct comparisons between the model runs which do and do not account for attractiveness and demand.

*Inferring flows from Twitter data*

Geographic tweets are discrete events that occur in continuous space and time with a high degree of random variability. Flow models, by contrast, use mathematics to describe smooth distributions over space and time. Even when flow models are stochastic, the underlying probabilities generally represent smooth distributions (38). In this context it is clear that spatial and temporal aggregation is needed to link Tweets with flow data. The temporal aggregation used in this example is simple: the sum of all tweets over the 445 days of collection. To coincide with the SIM, tweets were also aggregated spatially, allocated to a ward associated with the user's most frequent tweeting location and museums, one for each tweet sent. This process of spatial aggregation is a key step towards inferring flows from volunteered geographical data, and is shown in figure 5.

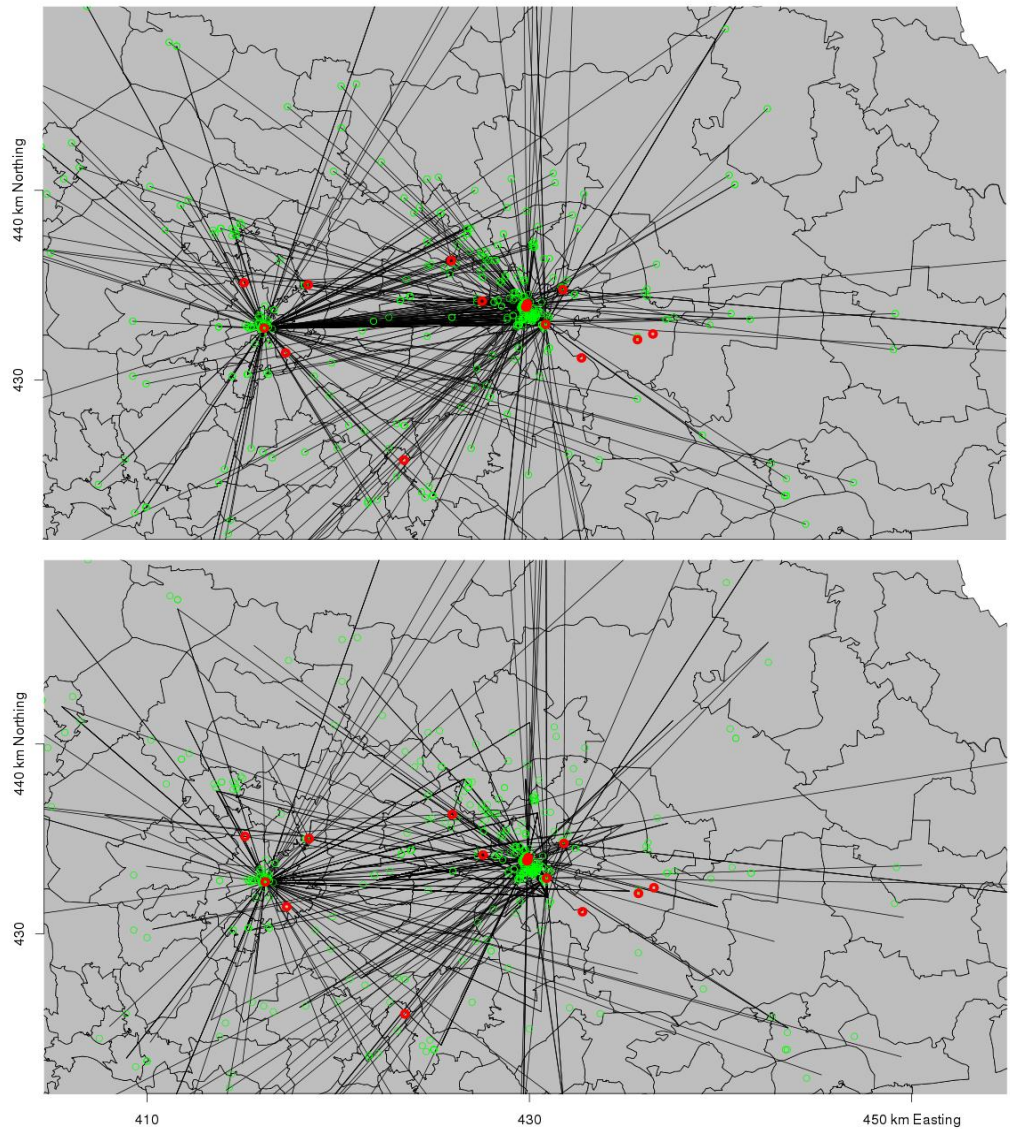

Figure 5. Flow maps of inferred museum visits from raw tweets (above) and from spatially aggregated tweet home locations (below).

*Comparing model output with tweets*

The process of spatial aggregation presented in Figure 5 is not only useful for visualizing the density of flows; it can also be used to generate an origin-destination flow matrix with the same dimensions as T in equation 4 for direct comparison with the SIM, which we shall refer to as T'. The pseudo code used to generate T' is presented

below:

```
T' <- matrix (0, nrow=nrow(S), ncol=ncol(S))   # initial conditions
for i in 1:n.tweets {
     T'[home[i], c.museum[i]] <- T'[home[i], c.museum[i]] + 1
}
```

Where home[i] represents the home ward of the person associated with tweet i and c.museum[i] is the index of the museum closest to the geographical coordinates of tweet i respectively. Thus T' is a sparse matrix containing positive integers only for the origin-destination flows represented in figure x and the sum of T' is equal to the total number of tweets (756 in this case). From this point comparison of T and T' can be achieved by concatenating the matrices into vectors and using standard measures of fit, such as Pearson's coefficient of correlation (r) and root mean squared (rms).

## Results: correlations and inconsistencies between Twitter data and model output

### *Baseline model*

The baseline scenario consisted of the simplest implementation of equation 4, with the variables Inc and W set to 1 for all wards and museums respectively. Thus, the only variable to optimize was $\beta$, which was initially set to 0.3, resulting in a positive correlation of 0.31 between the 2685 values of T and T'. Iterating through 200 model-tweet observation comparisons (step size = 0.01) it was found that model fit was optimal with a $\beta$ value of 0.95, with the correlation peaking with an r value of 0.39.

### *Including museum attractiveness*

The next model tested added the W variable – set by equation 6 – to the model, to make larger and more frequently mentioned museums more attractive than small museums that few people had heard of. The impact of this change was dramatic, with r values peaking at 0.60. The optimal distance decay function in this model specification was found to be steeper (β = 1.32).

*Including demand and attractiveness*

In the final model test the added refinement of variable demand from the origins was set, by altering Inc values according to equation 7. The performance of this model specification was found to be intermediate compared with the other two, with a maximum r value of 0.55. These results are displayed in figure 6 below.

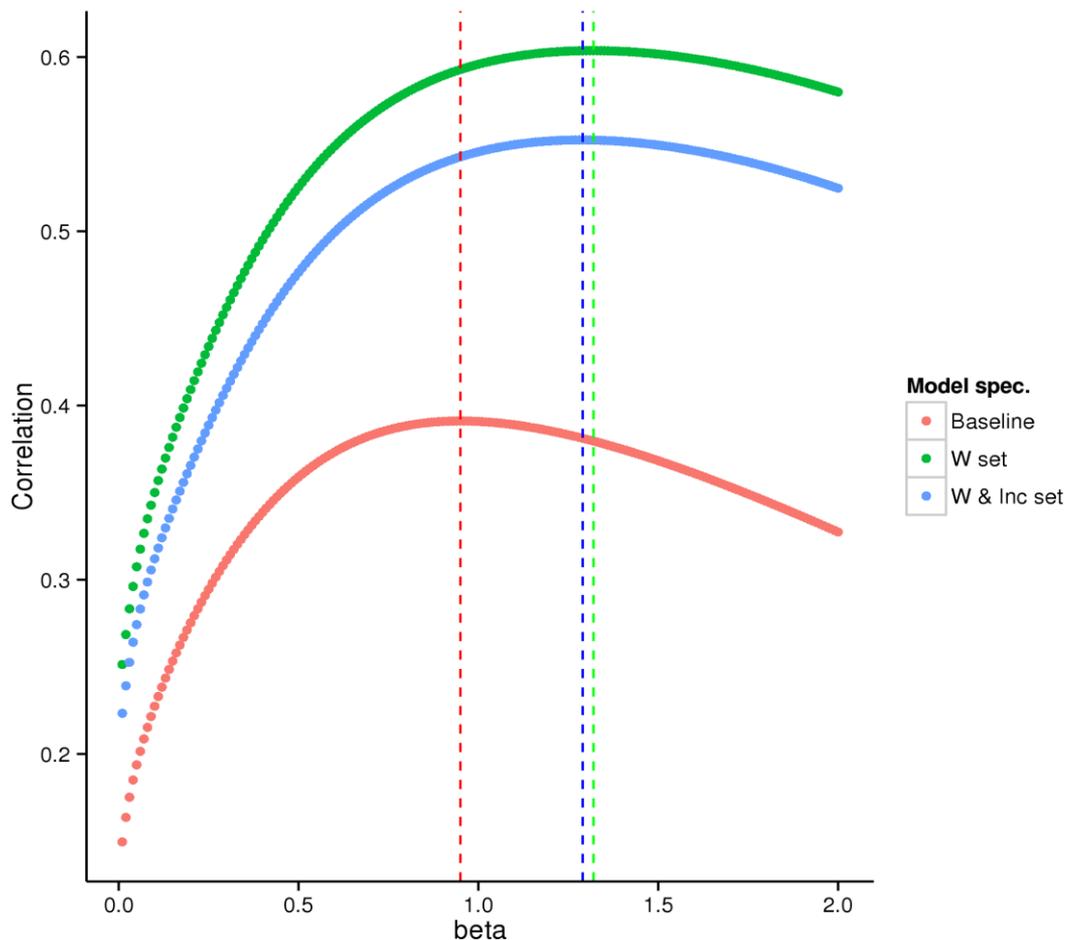

Figure 6. Correlation between the spatial interaction model (S) and flows inferred from tweets (S') against β values, for three different model specifications.

*Model enhancements*

There are many refinements that can be made to the basic models specified above. A major change would be to use a constrained spatial interaction model, whereby the flows from each origin are set (see figure 7). It was found that the model fit declined greatly in this scenario, however, due to forced flows to distant museums from wards in the periphery of the study region. The decrease in model fit in the constrained model fit was expected for two reasons:

- The study area is being treated as an encapsulated system which is artificially imposing a boundary that does not exist in the real world on the data. This has the effect of inflating flows originating near the edge of the study area where alternative attractions are being excluded because they fall on the wrong side of the boundary. This effect is more commonly known as the boundary effect.

- It is also likely that people living in remote areas simply visit museums less frequently due to accessibility.

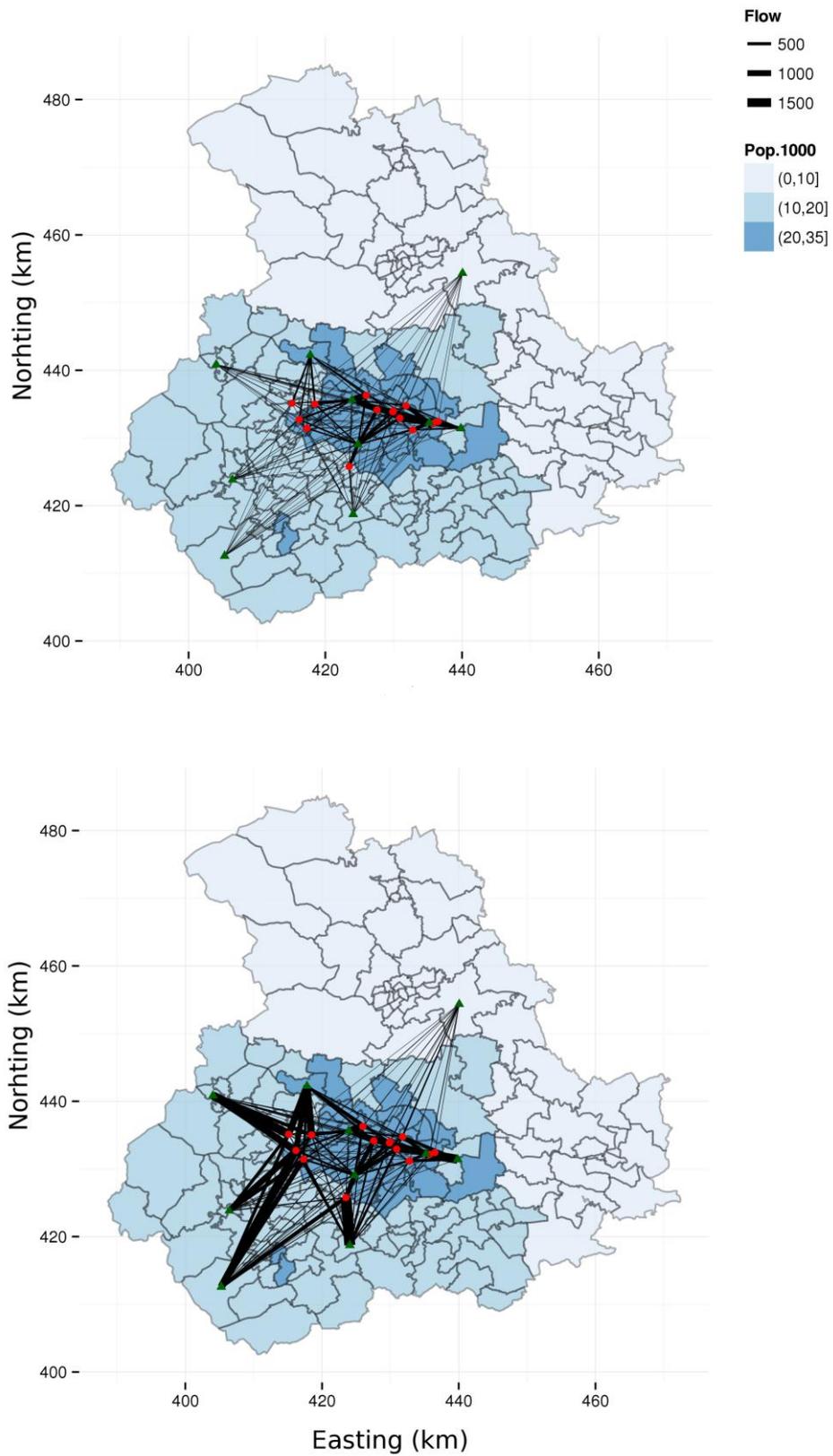

Figure 7. Constrained and unconstrained version of the model with W set. The green triangles are 20 randomly selected origins. Map projection: OSGB 1936.

A number of further refinements could be made to the estimation of demand and attractiveness, by further adjustments to equations 6 and 7. More experiments have been undertaken, but it is considered that the data is insufficiently large to reliably calibrate multiple parameters without undesirable 'over fitting' effects, whereby a model is essentially specified to fit noise within the data. Due to the relatively small sample size of museum tweets used in this study the model was designed simplistically and it was deemed unnecessary and potentially unsupportable to further complicate the model with additional parameters. Instead, the results form the basis of a discussion of the potential for VGI from social media to inform spatial interaction models in general terms. Based on this discussion, our methods can be seen as a stepping stone towards calibrating more complex models based on larger VGI datasets and more pressing real-world applications.

**Discussion and conclusion**

In this paper we have used geo-located tweets to calibrate a spatial interaction model representing visitor flows to museums from residential areas surrounding Leeds and Bradford in the UK. In the process we have highlighted the potential for VGI as an input into quantitative geographical research in general, and the use of Twitter data to calibrate SIMs in particular. Methods for filtering and aggregating the geo-located tweets have been developed, allowing the SIM to be calibrated by information made publicly available on the internet. Although the scope of the case study is limited geographically to one study area and semantically to museums, the study demonstrates the wider potential of VGI from social media in geographical research and provides a foundation for discussion of VGI in geo-science applications overall.

*Data limitations*

In agreement with Goodchild (3), we conclude that free, geographic and semantically rich datasets derived from the harvesting of social media sites en masse should continue to be of great and growing interest to all spatial analysts. Data quality remains a serious concern for all such VGI, however (2). The sheer diversity and sporadic nature of the data poses new challenges to researchers accustomed to relatively clean official datasets. As emphasized throughout, these challenges should not be overlooked: they must be acknowledged at the outset and tackled with care. More specifically, the main limitations of the data used in this study include:

- limited data availability

- the sub national nature of the study area

- a limited set of museums being considered, and

- uncertainty about the travel behaviour of those people living remotely to museum attractions.

Improved data harvesting and retrospective data collection could help overcome the first two issues; use of officially registered museums and visitor data could tackle the third and fourth points. The broader problem of wide semantic diversity can be partially tackled by intelligent filtering, using search terms that capture only messages that can be confidently attributed to the topic of interest. Nevertheless, there will inevitably still be variations in quality and issues of self-selection even after the most stringent filtering strategies. One way of tackling this issue would be to assign a continuous weight variable to every tweet corresponding to its relevance to the research problem, although the method of assigning high and low weights would add further complexity and subjectivity to the analysis. This point is related to a more fundamental problem with social media data: the context-dependence. Tweets sent to and from individuals contain many subtleties that are useful to users who can decode them. "Yet, taken out of context, data lose meaning and value" (39). In the process of harvesting and subsequent sampling much of this context is lost: currently there is no way to consider the wider context, implicit in each tweet, over thousands and indeed millions of such data points.

*Future development of VGI derived from social media*

The rapid penetration of mobile phones worldwide is being closely followed by uptake of smart phones, many of which contain GPS receivers and the capacity to interact with public social media websites such as Twitter, Facebook. Therefore the potential for growth of datasets such as the geo-tagged tweets presented in this paper is truly enormous, with equally huge potential for researchers (40). To some extent the aforementioned data limitations could be partly mitigated by sheer volume, increasing the signal-to-noise ratio (41). However, continued near-exponential future growth in publicly available geo-tagged social media is by no means certain: issues of data privacy and ownership could hamper the availability of these datasets to researchers working in the public domain.

There is a danger that companies such as Twitter restrict access to the data only to private companies with the financial resources to pay for the access. This is to some degree being mitigated by initiatives to make publicly available social media messages available through publicly owned centralized repositories such as the Library of Congress, which should allow for retrospective searches, rather than researchers' current reliance on real-time harvesting through APIs and retrospective scraping of data from the internet (42). Another issue is the transitory nature of social media websites: it has been hypothesized that popular social media sites such as Twitter and Facebook contain the seeds of their own demise, so it would be a mistake to assume uncritically that a single service can be relied upon for social VGI into the long-term. In addition increased public awareness of digital data archiving and analysis in the wake of leaked National Security Agency (NSA) documents could lead to a curtailment or alteration in the sharing of personal information such as location. Alternatively, it could be argued that making certain communications explicitly public, for social benefit, could become increasingly common as a counterbalance to the perceived concentration of digital intelligence gathering and processing capabilities by a few clandestine organizations.

*Future research potential*

Despite the data limitations, ethical concerns and unknown longevity of social VGI, it seems likely that the size and richness of available datasets will continue to grow. In parallel with this, computing power will continue to improve and computer programs will continue to develop towards greater functionality and user friendliness. This means VGI from social media will become an increasingly attractive alternative to official datasets for geographical problems that are presented in this paper, where data limitations remain a major constraint. There is clearly great potential for further research using this datasource in many areas, including the following:

- Geographically extensive events such as riots (43) and epidemics (44).

- The analysis of shifting attitudes and behaviors in relation to health or environmental drivers (45).

- The use of social-media VGI to inform and calibrate a-priori models of spatial behavior such as that presented in this paper.

Because the explosion of VGI is a recent phenomenon that challenges established habits of doing research, there is huge untapped potential in each of these areas and more. We urge researchers to think carefully about the research problems that they are interested in before evaluating whether or not geotagged social media is an appropriate input data source. In many cases it will not be: it is important that researchers do not simply follow the newest and largest datasets just because they are available and (potentially) free at source. It is hoped that this paper will lead to further discussion of the relative merits of Twitter and other volunteered social media information for informing geographical research. Ethical considerations should also guide the research: if the information is provided by the public for free, surely the benefits that accrue should be for public benefit.